\documentclass{nature}

\usepackage{url}
\usepackage[pdftex]{graphicx}

\title{Collective allocation of science funding: from funding agencies to scientific agency.}

\author
{Johan Bollen$^{1\ast}$, David Crandall$^{1}$, Damion Junk$^{1}$,
Ying Ding$^{2}$, and Katy B\"{o}rner$^{2}$
}

\begin{document}

\maketitle

\begin{affiliations}
\item School of Informatics and Computing, Indiana University, Bloomington IN 47401
\item School of Library and Information Science, Indiana University, Bloomington IN 47401
\end{affiliations}

\begin{abstract}






Public agencies like the U.S. National Science Foundation (NSF) and the National Institutes
of Health (NIH) award tens of billions of dollars in annual science
funding. How can this money be distributed as efficiently as possible
to best promote scientific innovation and productivity?  The present system
relies primarily on peer review of project proposals. In 2010 alone,
NSF convened more than 15,000 scientists to review 55,542
proposals\cite{nsfpeerreview}. Although considered the scientific gold
standard\cite{Jefferson2006}, peer review requires significant
overhead costs\cite{academiasuppresses,roy85}, and may be subject to biases,
inconsistencies, and
oversights\cite{Gura2002,subra:research2012,wessely98,peerreviewreliability,cole81consensus,nihreview,horrobin96,Editorial2013,JoshuaM.Ioannidis,smith06,marsh08,bornmann07,johnson08}.
We investigate a class of funding models in which all participants
receive an equal portion of yearly funding, but are then required to
anonymously donate a fraction of their funding to peers. The funding
thus flows from one participant to the next, each acting as if he or she
were a funding agency themselves.  Here we show through a simulation
conducted over large-scale citation data (37M articles, 770M citations)
that such a distributed
system for science may yield
funding patterns similar to
existing NIH and NSF distributions,
but may do so at much lower overhead while exhibiting a range of other
desirable features.  
Self-correcting mechanisms in scientific peer evaluation can yield an
efficient and fair distribution of funding.  The proposed model can be
applied in many situations in which top-down or bottom-up allocation of
public resources is either impractical or undesirable, e.g.  public
investments, distribution chains, and shared resource management.
\end{abstract}

Proposals to reform funding systems have ranged from incremental
changes to peer review including careful selection of
reviewers\cite{marsh08} and post-hoc normalization of
reviews\cite{johnson08}, to more radical proposals like opening up
proposal review to the entire online population\cite{calmcrisis} or
removing human reviewers altogether by allocating funds equally, randomly, 
or through an objective performance measure\cite{roy85}.  In contrast, we propose a
distributed system that incorporates the opinions of the entire
scientific community, but in a highly-structured framework that
encourages fairness, robustness, and efficiency.

In particular, in the proposed system all scientists are given an equal, fixed base amount of
yearly government funding. However, each year they are also required to distribute a given percentage of
their funding to other scientists whom they feel would make best use of
the money (Fig. 1). Each scientist thus receives funds
from two sources, the fixed based amount they receive from the government and the amount they receive from other scientists.
As scientists donate a fraction of their total funding to others scientists each year,
funding moves from one scientist to the next. Everyone is guaranteed the base amount, but  larger amounts 
will accumulate with scientists who are generally expected to make the best use of it.
 
For example, suppose the base funding amount is set to
\$100,000. This roughly corresponds to the entire NSF budget in 2010 divided by the total number
of senior researchers it funded\cite{nsfbudgetrequest}.
The required donation fraction is set to $F=0.5$. Scientist K
receives her yearly base amount of \$100,000, but in addition receives
\$200,000 from other scientists in 2012, bringing her total funding to \$300,000. 
In 2013, K may spend half of that total, i.e. \$150,000, on her own research program,
but must donate the other half to other scientists for their use in 2014.
Rather than painstakingly submitting project proposals, K and her colleagues only need to take a few minutes of their time
each year to log into  a centralized website and enter the names of the scientists they choose to
donate to and how much each should receive.

More formally, suppose that a funding agency maintains an account for
each of the $N$ qualified scientists (chosen according to some
criteria like academic appointment status or recent research
productivity), to whom we assign unique identifiers in $[1,N]$.  Let
$O_{i \rightarrow j}^t$ denote the amount of money that scientist $i$
gave to scientist $j$ in year $t$.  The amount of funding each
scientist receives in year $t$ is equal to the base funding $B$ from
the government plus the
contributions from other scientists,
\[ A^t_i = B + \sum_{j \in [1,N]} O_{j \rightarrow i}^{t-1}.\]
We require that every scientist gives away a fraction $F$ of their
funds each year,
\[ 
\sum_{j \in [1,N]} O^t_{i \rightarrow j} = (F) A^{t}_i \,\,\,\,\,\,\,\,\,\, \forall i \in N \]
while they are allowed to spend the remaining money on their research
activities.  Taken together, these two equations give a recursive
definition of the behavior of the overall funding system. A similar
formulation has been used to rank webpages by transferring influence
from one page to the next\cite{pagerank}, as well as to rank
scientific journals\cite{journa:bollen2006} and author
``prestige''\cite{Ding2009}.

This simple, highly distributed process yields surprisingly
sophisticated behavior at a global level.  First, respected and highly-productive
scientists are likely to receive a comparatively large number of
donations. They must in turn distribute a fraction of this total to others; their high 
status among scientists thus affords them both greater funding and greater influence over how funding is distributed. 
Second, as the priorities and preferences of the scientific community change over 
time, the flow of funding will change accordingly. Rather
than to converge on a stationary distribution, the system will dynamically adjust
funding levels as scientists collectively assess and re-assess each others' merits.

\section*{A large-scale simulation}

How would funding decisions made by this system compare to the gold standard of peer
review?  No funding system will be optimal since research outcomes
and impact are difficult to predict in advance\cite{myhrvold98}. At the very least 
one would hope that the results of the proposed system match those of existing funding systems.

To determine whether this minimal criterion is satisfied, we conducted a large-scale,
agent-based simulation to test how the proposed funding system might operate in practice.
For this simulation we used citations as a proxy for how each scientist might
distribute funds in the proposed system.
We extracted 37 million academic papers and their 770 million references from
Thomson-Reuters' 1992 to 2010 Web of Science (WoS) database. We
converted this data into a citation graph by matching each reference
with a paper in the database by comparing year, volume, page number,
and journal name, while allowing for some variation in journal names
due to abbreviations. About 70\% of all references could be matched back 
to a manuscript within the WoS data itself.

From the matching 37 million papers, we retrieved 4,195,734
unique author names; we took the 867,872 names who had authored at least one
paper per year in any five years of the period 2000--2010 to be in the
set of qualified scientists for the purpose of our study.
For each pair of authors we determined the number
of times one had cited the other in each year of our citation data (1992--2010).
We also retrieved NIH and NSF funding records
for the same period, a data set which provided 347,364 grant amounts for 
109,919 unique scientists\cite{larowe07}.

We then ran our simulation beginning in the year 2000, in which everyone
was given a fixed budget of $B=\$100,000$.  We simulated the
system by assuming that all scientists would distribute their funding
in proportion to their citations over the prior 5 years. For example,
if a scientist cited paper A three times and paper B two times over the
last five years, then she would distribute three-fifths of her budget
equally among the authors of A, and two-fifths amongst the authors of B. 
  
Note that we are merely using
citation data in our simulation as a proxy for how scientists \textit{might} distribute their funding;
we are not proposing that funding decisions be made on the basis of citation analysis.

The simulation  suggests that the resulting funding
distribution would be heavy-tailed (Fig. 2a), similar in shape to 
the actual funding distribution of NSF and NIH for the period 2000-2010. As expected,
the redistribution fraction $F$ controls the 
shape of the distribution, with low values creating a nearly uniform
distribution and high values creating a highly biased distribution.

Finally, we used a very conservative (and simple)
heuristic to match the author names from our simulation
results to those listed in the actual NSF and NIH funding records: we
simply normalized all names to consist of a last name, a first name, and
middle initials, and then required exact matches between the two
sets. This conservative heuristic yielded 65,610 matching scientist
names. For each scientist we compared their actual NSF and NIH funding for 2000--2010
to the amount of funding predicted by simulation of the
proposed system, and found that the two were correlated with Pearson
$R=0.268$ and Spearman $\rho=0.300$ (Fig. 2b).

\section*{Discussion and conclusion}

These results suggest that our proposed system would lead to
funding distributions that are highly similar in shape and individual level of funding
to those of the NIH and NSF, if scientists are compelled to redistribute 50\% of
their funding each year, but at a fraction of the time and cost
of the current peer-review system.

Note that instead of funding \emph{projects}, the proposed framework would fund \emph{people}\cite{Gilbert2009}. This gives
scientists more freedom and opportunity for serendipitous discovery, rather than to chase available
funding\cite{Editor2011}.  Of course, funding agencies and governments may still wish to play a directive role, e.g.
to foster advances in certain areas of national interest or to foster diversity.
This could be included in the outlined system in 
a number of straightforward ways. Most simply, traditional peer-reviewed, project-based funding
could be continued in parallel, using our system to distribute a portion of
public science funding.  Traditional avenues may also be needed
to fund projects that develop or rely on large scientific tools and infrastructure. Funding
agencies could vary the base funding rate $B$ across different
scientists, allowing them to temporarily inject more money into certain disciplines.
The system could also include some explicit temporal dampening to
prevent large funding changes from year to year.

In practice,  the system may also require conflict of interest rules similar 
to the ones used to keep traditional peer-review fair and unbiased.
For example, scientists might be prevented
from contributing to advisors, advisees, close collaborators, or
researchers at their own institution.  Scientists might
also be required to keep their funding decisions confidential in order to
prevent groups of people from colluding to affect the global funding
distribution.

Peer-review of funding proposals has served science well for decades,
but funding agencies may want to consider alternative approaches to
public science funding that build on theoretical advances and leverage
modern technology to optimize return on investment.  The system
proposed here requires a fraction of the costs associated with peer
review, but may yield comparable results. The savings of financial and
human resources could be used to identify targets of opportunity, to
translate scientific results into products and jobs, and to help
communicate advances in science and technology to a general audience
and Congress.

\section*{References}
\bibliography{fundrank_arxiv}

\begin{addendum}
 \item The authors acknowledge the generous support of the National Science Foundation under grant SBE \#0914939 and the Andrew W. Mellon Foundation. We also thank the Los Alamos National Labotory Research Library, the LANL  Digital Library Prototyping and Research Team, Thomson-Reuters, and the  Cyberinfrastructure for Network Science Center at Indiana University for furnishing the data employed in this analysis.

 \item[Competing Interests] The authors declare that they have no
competing financial interests.
 \item[Correspondence] Correspondence and requests for materials
should be addressed to Johan Bollen~(email: jbollen@indiana.edu).
\end{addendum}

\section*{Figures and tables}

\includegraphics[width=\textwidth]{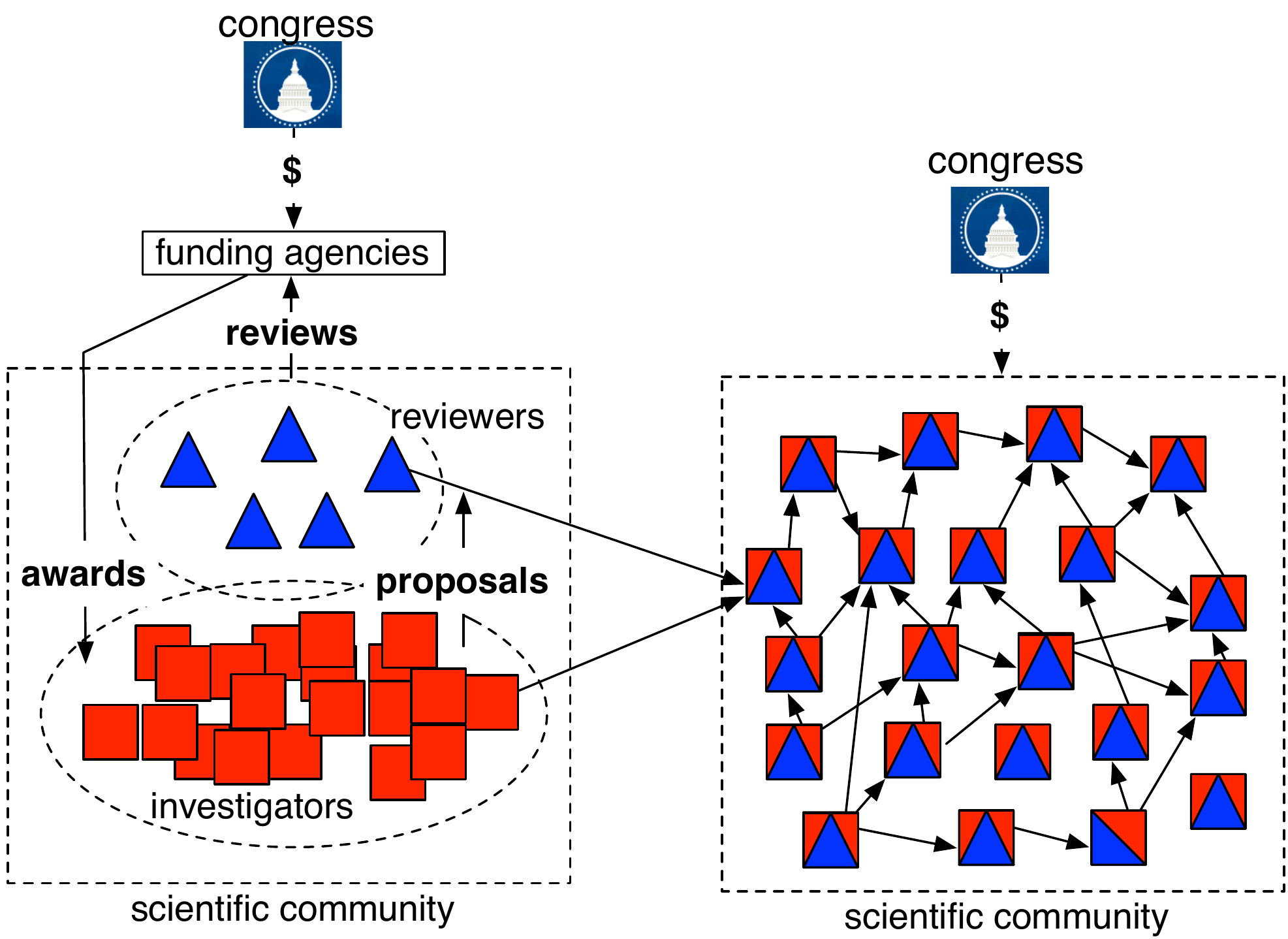} 
\label{fig:diagram}
\hspace{0.6in} \textbf{Existing funding system} \hspace{1.6in} \textbf{Proposed funding system} \\
\textbf{Fig. 1:} Illustrations of existing (left) and the proposed (right)
  funding systems, with reviewers marked with triangles and investigators imarked by squares. In
  most current funding models like those used by NSF and NIH,
  investigators write proposals in response to solicitations from
  funding agencies, these proposals are reviewed by small panels, and
  funding agencies use these reviews to help make funding decisions,
  providing awards to some investigators. In the proposed system all
  scientists are \emph{both} investigators and reviewers: every scientist
  receives a fixed amount of funding from the government and other scientists but is
  required to redistribute some fraction of it to other investigators.

\includegraphics[width=0.48\textwidth]{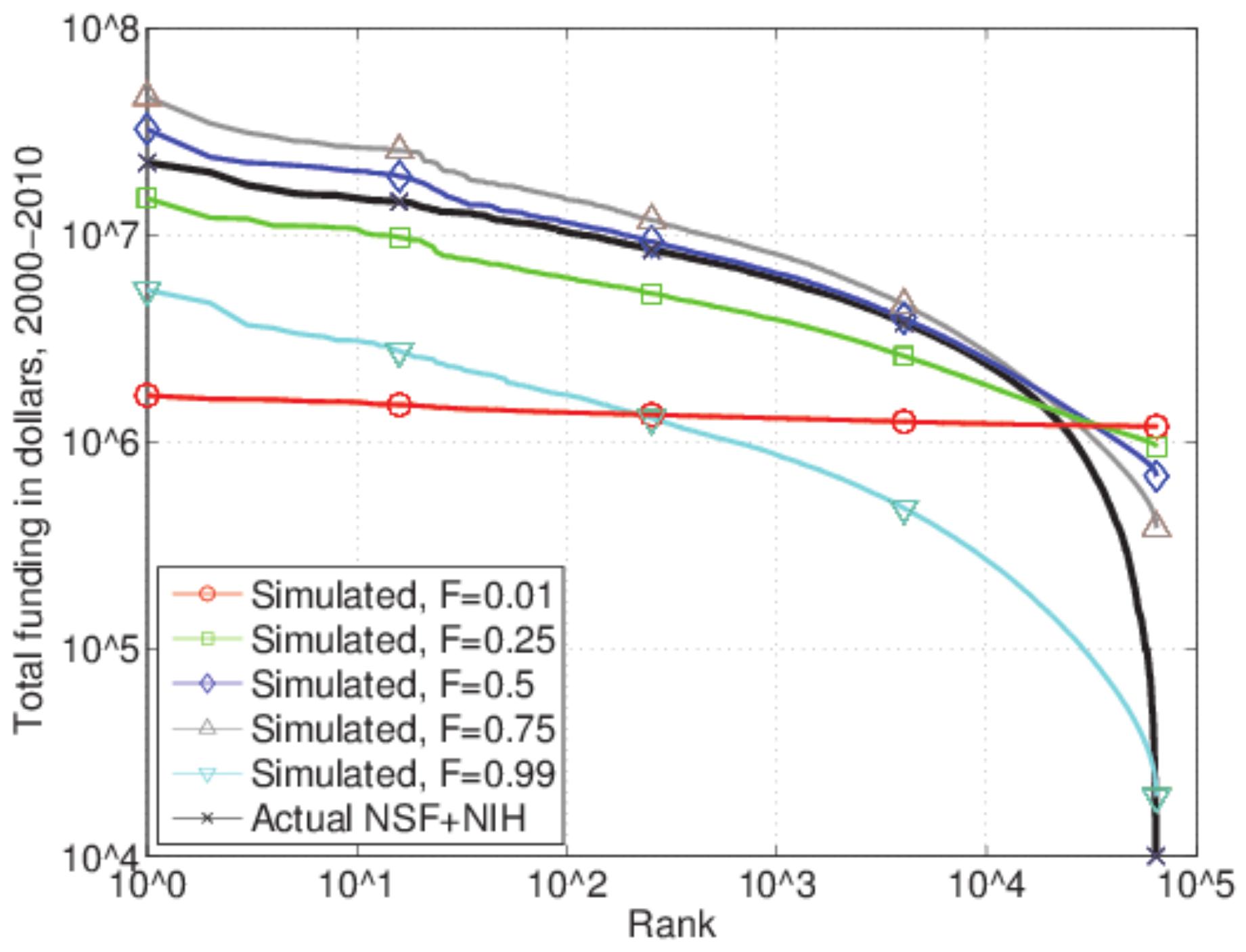}
\includegraphics[width=0.48\textwidth]{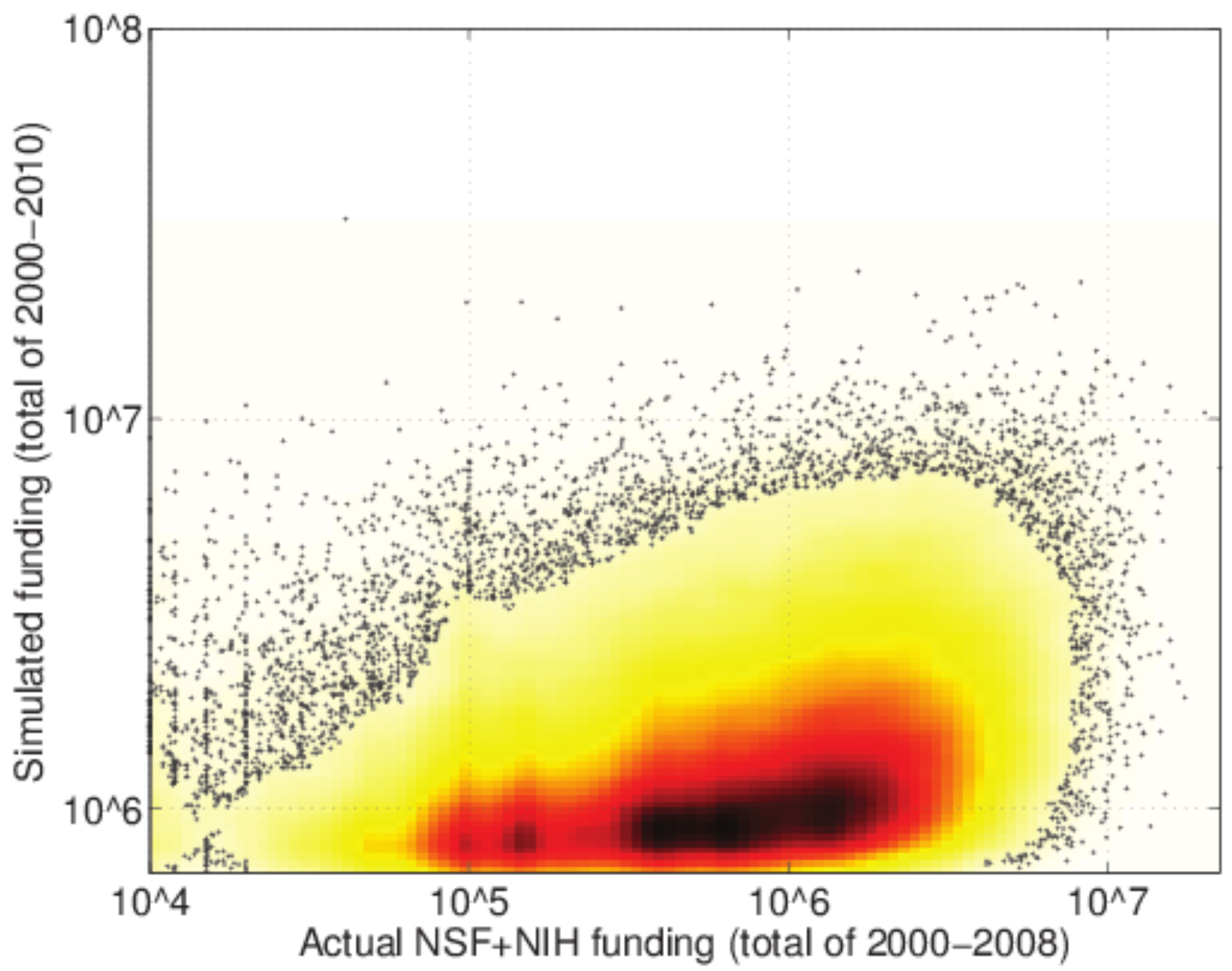}
\label{fig:plots}
\textbf{Fig. 2:} Results of the distributed funding system simulation for
  2000-2010.  \textit{(a):} The general shape of the funding distribution is similar to that of  actual
historical NSF and NIH funding distribution. The shape of the distribution can be
controlled by adjusting $F$, the fraction of funds that scientists
must give away each year. \textit{(b):} On a per-scientist basis, simulated funding from our system
(with $F$=0.5) is correlated with actual NSF and NIH funding (Pearson $R=0.2683$ and Spearman $\rho=0.2999$).

\newpage

\section*{Supplemental materials}

\textbf{Overview}

The simulation of our proposed funding system (which we call FundRank) was based on the assumption that we could use authors' citation behavior
as a proxy of their potential donation behavior. In other words, we assumed that we could estimate
\textit{whom people would donate funding to} based on \textit{whom they frequently cited in the recent past}.

To determine author citation behavior we created an author-to-author citation network from a large-scale source
of article citation data as follows:

\begin{enumerate}
	\item Extract an article-to-article citation network from 20 years of Thomson-Reuters' 
		Web of Science (WoS) reference data;
	\item Extract the authors from each of the articles in our article citation network;
	\item Aggregate article-to-article citations into author-to-author citations;
	\item Created an author citation network for each year of the mentioned 20 years of WoS data.
\end{enumerate}

\textbf{Data}

This analysis is based on Web of Science (WoS) citation data that was kindly made
available to our project by Thomson-Reuters, by way of the Los Alamos National Laboratory
(LANL) Research Library (RL), where it was pre-processed by the Digital Library Prototyping
and Research Team of the LANL RL (please see acknowledgements).

Our WoS data spanned 20 years (1990 to 2010) and offered bibliographic data for a total of
37.5M publications. Each primary bibliographic record in the data corresponded to one unique
scholarly publication for which the record provided a unique identifier, the publication date, issue,
volume, keywords, page range, journal title, article title, volume,  and the record's bibliography (list
of references).

The references indicate which articles are cited by the primary record, but 
consisted of a summarized citation that only contained a single author, year, page number, journal title,
and volume. Not all references contained values for each of the mentioned fields, and no unique
identifiers were provided. A total of about 770M reference records were contained within the 37.5M
primary bibliographic records in our data.

The primary records and their references should in principle map to the same set of articles, establishing
a citation relation between the primary bibliographic record and the articles it references, but given the
differences in formatting and the lack of bibliographic information in the references 
our set of bibliographic records was thus initially 
separated into two separate types of data: (1) 37.5M primary records, and (2) a total of about 770 million
summary reference entries contained by the former. 


\textbf{Reference metadata matching for article citation network creation}

To establish an article citation network over the 37.5M primary records in our data, it was
necessary to determine which of the 770M million reference entries mapped back to any of the 
37.5 primary bibliographic records. In other words, we had to determine whether an 
article A cited another article B by determining whether any of A's references matched the bibliographic
information of article B.
The reference data contained only abbreviated journal titles and abbreviated author names 
(typically only the first author), so we needed to match this information to the more detailed 
bibliographic data provided in the primary records in the WoS data.  
To do this, we assigned a metadata-based identifier to each primary record,
$$
\mbox{ID} = (\mbox{journal name}, \mbox{journal volume}, \mbox{page number}, \mbox{year of publication}),
$$
which we expected to provide a reasonably unique identifier since it consisted of bibliographic information that was
1) available for both primary records and references, and 2) well-defined, unambiguous numerical information.

\includegraphics[width=12cm]{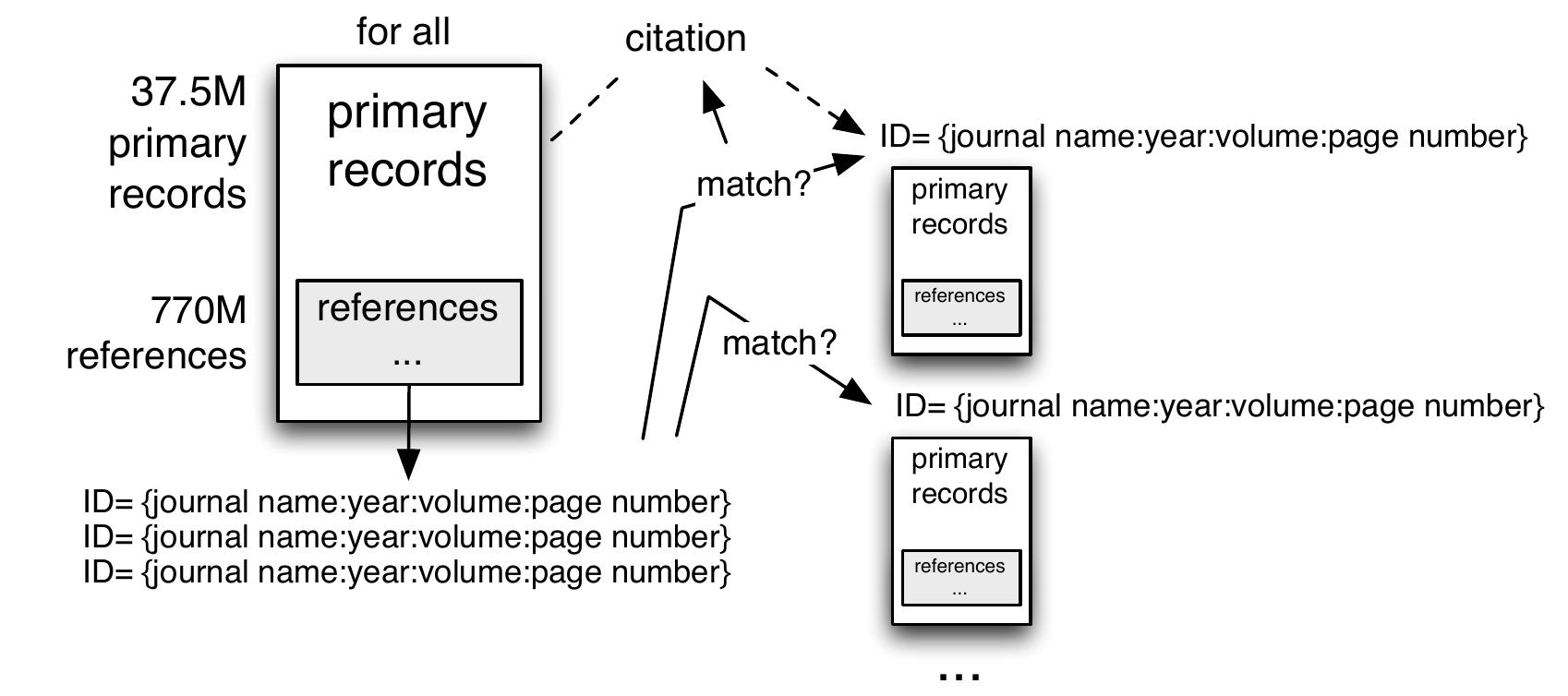}\\
\label{record_reference_matching}
\textbf{Fig. 3:} Matching of metadata-based identifiers generated for (1) references to those generated for (2) primary records to determine citation relations between the 2 primary records involved.

As shown in Fig. 3, we then attempted to match the article identifier generated for all 37.5M primary bibliographic records to those of all 770M references. Each metadata identifier between reference vs. primary record match was taken to indicate a citation relation between the matching primary records.

In doing this matching, we allowed for imperfect, partial matches on
some elements of the metadata identifier to account for errors and typos
and references:

\begin{description}
\item[-- Page numbers:] The page number could be either within the range indicated by the master, e.g. ``1540'' matches ``1539-1560,''
or an exact text match of the page entry for the primary and reference identifier, e.g ``13a.''
\item[-- Journal titles:] Due to significant variation of journal names, for example differing 
and inconsistent abbreviations,  partial matches were allowed for journal titles.
\end{description}

For this latter item, we defined a heuristic algorithm to detect
matching journal titles across various spelling and abbreviation
standards, as shown in Fig. 4.

\includegraphics[width=14cm]{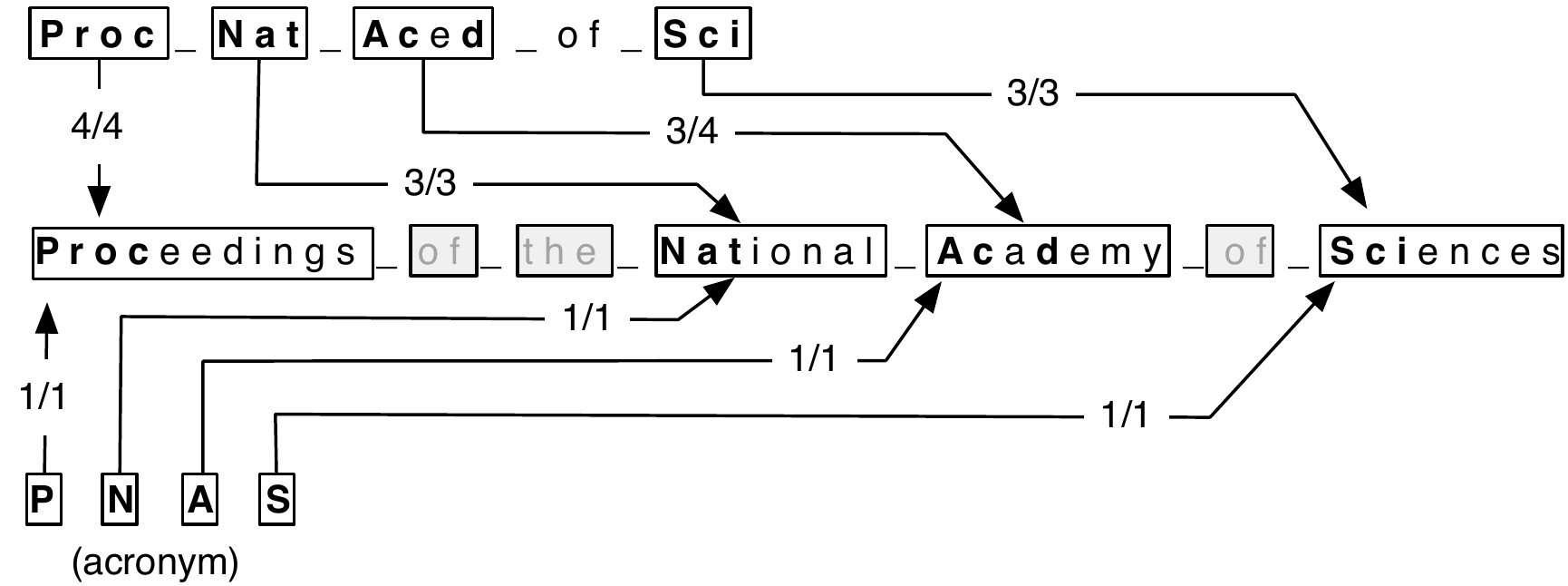}\\

\textbf{Fig. 4:} Matching abbreviated journal title
  variants by means of longest common substring matching, expanded to
  handle acronyms.

 First, numbers, symbols, and stop words were removed, and
repeated spaces were reduced to a single space. All characters were
converted to lower-case.  Second, the resulting titles were split into
individual terms, which were scanned character by character from left
to right to compute the degree of overlap between the individual
space-delimited words of each title.  Then, for each pair of terms
across the titles, we calculated the fraction of the  characters
in the shortest term that matched the characters of the longest term
without interruption from the left to the right over the length of the
shortest term.

For example, the two titles "Proc Nat Acad Sci" and "proceedings of
the national academy of sciences" would first have numbers, symbols,
and stop words removed, after which each pair of terms in the
resulting titles would be compared character by character to determine
their degree of overlap. That is, \textit{``proc''} would be compared
to \textit{``proceedings''}, \textit{``nat''} to
\textit{``national''}, etc.  All 4 characters of ``acad'' match the
first 4 characters of ``academy'', and therefore the two terms were
considered a perfect match.  The average of the ratios across the
entire titles produced a ``match ratio'' which
could be used to assess the degree to which they referred to the same
journal.

We included an additional heuristic to handle journal title acronyms.
If a title was less than six characters in length and contained no
spaces it was considered an acronym. In our matching system, each
letter of the acronym was considered as an individual word to be
matched against possible targets in the longer title. This allows for
comparison such as ``PNAS'' vs. ``Proc. Nat. Acad. Sci'', and ``PNAS''
vs. ``Proceedings of the National Academy of Sciences'' to result in
positive matches.

Finally, a reference record was considered a match when the (year of
publication, journal volume, page number) tuple matched exactly, and
the journal title match score exceeded a configurable threshold; we used
90\% in this paper.

This procedure for matching reference identifiers to primary record identifiers
allowed us to connect nearly 70\% of all references to a primary
record, thereby achieving an article to article citation network with
wide coverage across the entire set of 37.5M primary records and 20
years of our WoS data.

\textbf{Author to Author Edge Lists}

In the 37 million papers extracted from the WoS data, we found
4,195,734 unique author names.  In principle it is trivial to derive
an author-to-author network from the established article-to-article
citation network by simply looking up the author names of the primary
records, and collating citation numbers across all publication records
of the individual authors. Because we were interested activity over a
6 year sliding window, only citations within that window are
considered when building our author-to-author network.

Unfortunately, in practice this procedure is difficult because there
is not a one-to-one mapping between author names and scientist names,
because (a) different scientists may have the same name, (b) some
scientists publish under different names (most notably inconsistently
using middle names or middle initials, or using nicknames), and (c)
there are typos and other errors in the WoS data.

We first attempted to collapse together duplicate names for the same
author.  How best to do this depends on the name in question; for a
very unique last name, for example, it is sensible to collapse more
aggressively than for common names. We used the following heuristic to
do this.  We partition the list of approximately 4 million names into groups
of mostly-similar names, i.e. where the last name and first initial
are the same, and proceeded to apply the following procedure to each group:

\begin{enumerate}
  \item If there's only a single name in the group, then we're done.
  \item If there are multiple names, then look at each of them in sequence. For each name X:
    \begin{enumerate}
    \item test whether X is a ``subname'' of exactly one other name in the
      group, i.e. an abbreviated form of the same name. For example: \\
      
	\texttt{Lastname, D J } is a subname of \texttt{Lastname, Daniel J } \\
	\texttt{Lastname, Daniel }  is a subname of  \texttt{Lastname, D J } \\
	\texttt{Lastname, D } is a subname of \texttt{Lastname, D J } \\
	\texttt{Lastname, Dan } is a subname of \texttt{Lastname, Daniel } \\

      If so, then merge those two names together.
    \item if X is a subname of multiple other names in the group (a set
      Y of names), then look at all of the names in {X} and Y. If they are all
      ``mutually compatible'' with one another, meaning that they all could
      refer to the same person, then merge them all together. If not, then
      do nothing because there is an inherent ambiguity that can't be
      resolved without additional information.
      
    \end{enumerate}
\end{enumerate}

This procedures amounts to a rather aggressive approach to collapsing names, but it 
stops whenever ambiguity arises. For example, if the set of author names 
includes ``Lastname, David'', ``Lastname, D'',
``Lastname David J'', and ``Lastname D J'', then they are all collapsed
into a single author. However if there are also some additional names
like ``Lastname, Daniel'', ``Lastname, Dan'', ``Lastname, Donald'',
``Lastname, D X'', and ``Lastname, Donald X'', then we end up with the
following collapsed equivalence classes:

\begin{verbatim}
	1: {"Lastname, David", "Lastname, David J", "Lastname, D J"}
	2: {"Lastname, Daniel", "Lastname Dan"}
	3: {"Lastname, Donald", "Lastname D X", "Lastname Donald X"}
	4: {"Lastname, D"}
\end{verbatim}
The last one becomes a singleton author set because we simply can not
resolve the ambiguity without more information.

\textbf{FundRank simulation}

From the resulting list of unique scientists, 
we filtered out people who had
not authored at  least one paper per year in any five years of the period 2000-2010.
The remaining 867,872 are the group of authors for
whom we conduct our FundRank simulation (our Scientists).

The FundRank simulation is carried out as follows.  On Jan 1 of each
year, each Scientist receives \$100,000 as their equal share of the
total amount of available funding.  On Dec 31 of each year, all
Scientists must donate a fraction $F$ of their funding to others,
distributed according to the number of citations that points from
their papers written that year to other authors, with the restrictions
that (a) Scientists cannot contribute money to themselves (even if
they cited their own paper) and (b) papers that are more than 5 years
old do not count.

In other words, if an author cited $n$ papers in a given year, and each one had
an average of $m$ authors per paper, then this author splits his
contributions across the $mn$ (not necessarily distinct) Scientists.
If a person didn't write any papers in a given year, or didn't cite
any papers, they simply distribute their money uniformly across the
entire community of Scientists.

\textbf{Correlations with NSF/NIH funding}

We received NIH and NSF funding data from the Cyberinfrastructure for
Network Science Center at the School of Library and Information
Science at Indiana University. The NIH data lists all details
pertinent to 451,188 grants that were made to Principal Investigators
(PIs) from January 1990 through the end of 2011. For each grant, the
dataset includes the PI name, the award date, PI institution, award
amount in US Dollars, and the grant's subject keywords.  The NSF data
lists all details pertinent to 198,698 grants awarded from 1990-2011:
PI name, award date, PI institution, award amount, and NSF program.
Both datasets only list the number of co-PIs but do not include their
names, so all comparisons are performed for the set of PIs listed
only.

Many grants are quite small, and intended to fund small workshops and teaching
needs instead of research projects. We attempted to remove these by
filtering out any awards of less than \$2,000 USD. Similarly, a few
awards are unusually large, and correspond to multi-institution grants for major
equipment development (e.g. building telescopes) that would be outside
the scope of FundRank. We thus also filtered out single awards greater
than \$2M USD as well.

We used a very conservative (and simple) heuristic to match up PI
names between the NSF and NIH datasets and the author names from our
FundRank simulation results: we simply normalized all names to consist
of a last name and first and middle initials, and then required \emph{exact}
matches between the two sets. This conservative heuristic yielded
65,610 matching author/PI names, which were then used to perform
correlations.

\end{document}